\documentclass[conference]{IEEEtran}
\IEEEoverridecommandlockouts

\usepackage{cite}
\usepackage{amsmath,amssymb,amsfonts}
\usepackage{algorithmic}
\usepackage{graphicx}
\usepackage{textcomp}
\usepackage{xcolor}
\def\BibTeX{{\rm B\kern-.05em{\sc i\kern-.025em b}\kern-.08em
    T\kern-.1667em\lower.7ex\hbox{E}\kern-.125emX}}

\usepackage{booktabs}        
\usepackage{threeparttable}  
\usepackage{siunitx}        
\usepackage[caption=false,font=footnotesize]{subfig}
\usepackage{tikz}
\usepackage{pgfplots}
\pgfplotsset{compat=1.18}
\usetikzlibrary{patterns}
\usepackage[table]{xcolor}
\usepackage{colortbl}

\newcommand{\ShadeCap}{60}
\newcommand{\scaleShade}[1]{\the\numexpr #1*\ShadeCap/100\relax}

\newcommand{\heatTeal}[2]{\cellcolor{teal!\scaleShade{#1}}{#2}}
\newcommand{\heatBlue}[2]{\cellcolor{blue!\scaleShade{#1}}{#2}}
\newcommand{\heatOrange}[2]{\cellcolor{orange!\scaleShade{#1}}{#2}}

\begin{document}

\title{Measuring Memecoin Fragility

}

\author{
Yuexin Xiang$^{1}$,
Qishuang Fu$^{1}$,
Yuquan Li$^{2}$,
Qin Wang$^{3}$,
Tsz Hon Yuen$^{1}$,
Jiangshan Yu$^{4}$
\\[1ex]
\IEEEauthorblockA{$^{1}$ Faculty of Information Technology, Monash University, Melbourne, Australia}
\IEEEauthorblockA{$^{2}$ Faculty of Business and Economics, The University of Melbourne, Melbourne, Australia}
\IEEEauthorblockA{$^{3}$ CSIRO Data61, Sydney, Australia}
\IEEEauthorblockA{$^{4}$ School of Computer Science, The University of Sydney, Sydney, Australia}
}


\maketitle

\begin{abstract}
Memecoins, emerging from internet culture and community-driven narratives, have rapidly evolved into a unique class of crypto assets. Unlike technology-driven cryptocurrencies, their market dynamics are primarily shaped by viral social media diffusion, celebrity influence, and speculative capital inflows. 

To capture the distinctive vulnerabilities of these ecosystems, we present the first \textit{Memecoin Ecosystem Fragility Framework} (ME2F). ME2F formalizes memecoin risks in three dimensions: \textit{i)} Volatility Dynamics Score capturing persistent and extreme price swings together with spillover from base chains; \textit{ii)} Whale Dominance Score quantifying ownership concentration among top holders; and \textit{iii)} Sentiment Amplification Score measuring the impact of attention-driven shocks on market stability. 

We apply ME2F to representative tokens (over 65\% market share) and show that fragility is \underline{not} evenly distributed across the ecosystem. Politically themed tokens such as TRUMP, MELANIA, and LIBRA concentrate the highest risks, combining volatility, ownership concentration, and sensitivity to sentiment shocks. Established memecoins such as DOGE, SHIB, and PEPE fall into an intermediate range. Benchmark tokens ETH and SOL remain consistently resilient due to deeper liquidity and institutional participation. Our findings provide the first ecosystem-level evidence of memecoin fragility and highlight governance implications for enhancing market resilience in the Web3 era.
\end{abstract}


\section{Introduction}

Memecoins, crypto-assets born from internet memes and propelled by online communities~\cite{shifman2013memes}, have emerged as a distinctive market segment within the broader Web3~\cite{conlon2025memecoin}. In contrast to technology-driven cryptocurrencies with intrinsic functionality, the value of memecoins is largely narrative-driven, shaped by social diffusion, celebrity endorsements, and speculative flows~\cite{kalacheva2025detecting}. Tokens such as Dogecoin (DOGE), Shiba Inu (SHIB), PEPE, and Official Trump (TRUMP) illustrate this dynamic, where community sentiment and virality dominate participation and valuation. As of October 2025, the memecoin market has reached a total capitalization of approximately USD 84 billion (abbreviated as \$84B)~\cite{coingecko}, highlighting its emergence as a distinct and sizable segment within the cryptocurrency ecosystem. 

However, reliance on narratives rather than fundamentals also makes these assets particularly vulnerable to structural risks~\cite{taffler2024narrative}. 
Because memecoins lack strong intrinsic utility, their prices become disproportionately exposed to behavioral shocks, coordination frictions, and targeted manipulation. For instance, TRUMP surged sharply after its launch in early 2025 and then lost over 80\% of its value within months, reflecting the characteristic of a rapid boom-and-bust dynamic.  

Unfortunately, existing research remains fragmented, often limited to single-asset studies of volatility or attention~\cite{nani2022doge,chen2025fragility}. The broader ecosystem-level fragility across multiple blockchains has not been consistently measured. Moreover, weaknesses in governance, such as opaque token allocations, concentrated ownership, and thin liquidity, exacerbate these vulnerabilities  and amplify cyclical instability.

To address these gaps, we conduct the first ecosystem-level empirical study of major memecoins, systematically quantifying fragility through the \textit{\underline{M}emecoin \underline{E}cosystem \underline{F}ragility \underline{F}ramework} (ME2F). Our framework defines three complementary scores: \textit{Volatility Dynamics Score} (VDS), capturing persistent and extreme fluctuations with adjustments for market scale and base-chain spillovers; \textit{Whale Dominance Score} (WDS), assessing concentration through both cumulative top-holder shares and their internal inequality; and \textit{Sentiment Amplification Score} (SAS), evaluating how underlying sentiment instability translates into amplified price reactions to external shocks. By operationalizing fragility through these scores, ME2F provides the basis for our empirical analysis.

\smallskip
\noindent\textbf{Contributions.} This paper makes three contributions to the study of memecoin ecosystems:
\begin{itemize}
    
\item We propose a new measuring framework ME2F to assess memecoin fragility across three dimensions: volatility dynamics, whale dominance, and sentiment amplification. For each dimension, we develop quantitative indicators by defining specialised scores, enabling systematic measurement and cross-asset comparison of memecoin fragility.

\item We apply the framework to seven major memecoins covering over 65\% of the market to evaluate volatility, concentration, and sentiment effects. The analysis reveals highly centralized ownership structures, with the top 100 addresses typically holding more than 70\% and in some cases exceeding 90\% of the total supply, together with pronounced volatility clustering and strong sentiment-price comovement.

\item We further discuss ecosystem-level fragility revealed by cross-metric analysis, showing that volatility, sentiment amplification, and ownership concentration constitute distinct yet overlapping sources of vulnerability across token categories. These insights inform governance design and regulatory oversight, underscoring the heightened exposure of politically themed and community-driven memecoins relative to more resilient base-layer assets.
\end{itemize}

\section{Memecoin Research Landscape}

\subsection{Overview of the Memecoin Ecosystem}

Memecoins, though lacking a formal definition, broadly refer to crypto-assets inspired by internet memes and sustained by online communities. They encompass independent tokens, wrapped assets, and forks with diverse governance and cultural framings. According to Coingecko~\cite{coingecko}, more than 5,600 memecoins have been deployed across various chains, yet only a small share sustain long-term liquidity or user participation. This heterogeneity makes memecoins a distinctive, rapidly evolving segment of the Web3 economy.

Building on prior research, we identify three recurrent characteristics that define memecoin behavior: 
(i) community-driven volatility and spillovers, 
(ii) polarized outcomes and whale dominance, and (iii) narrative-based valuation and affective utility. 
These characteristics jointly capture the interaction between financial volatility, social diffusion, and governance concentration in memecoin ecosystems.

\subsection{Feature I: Community-driven Volatility and Spillovers}

Memecoins exhibit community-driven volatility and cross-market contagion shaped by speculative coordination rather than fundamentals.  
During the pandemic, volatilities and correlations rose across Bitcoin (BTC), Ethereum (ETH), DOGE, and gold, with DOGE behavior consistent with fear-of-missing-out (FOMO) dynamics~\cite{zhang2021popular}.  
Connectedness analyses reveal recurrent bubbles and co-explosivity between meme stocks and cryptocurrencies~\cite{aloosh2022bubbles}, showing that meme assets amplify contagion in bullish phases but absorb shocks during downturns~\cite{yousaf2023connectedness}.  

At the platform level, rapid memecoin growth boosts revenues while heightening systemic risk and leaving political tokens especially exposed to contagion~\cite{conlon2025memecoin}.  
Complementary studies using cross-quantilograms find that Bitcoin provides limited safe-haven protection for newer assets, including memecoins, under volatility and sentiment extremes~\cite{chopra2024does}.  
Spillover tests further show bidirectional transmission: large-cap coins usually lead memecoins, yet positive net spillovers from memecoins can precede market-wide corrections~\cite{li2022will, li2023can}.  
Taken together, these studies portray memecoins as volatility amplifiers where online coordination and speculative sentiment jointly drive systemic interconnectedness.

\subsection{Feature II: Polarized Outcomes and Whale Dominance}

A second defining feature is extreme payoff polarization coupled with severe ownership concentration.  
Mongardini et al.~\cite{mongardini2025midsummer} find that over half of high-return tokens display alarming centralization, with the top ten holders often controlling 70--80\% of supply.  
Krause~\cite{krause2025meme} identifies even greater concentration in politically themed tokens such as TRUMP and MELANIA, where single wallets initially held over 80--90\%, enabling coordinated liquidation and insider profit extraction.  

On-chain audits show that over 98\% of newly minted tokens on Ethereum and Solana display rug-pull or bot-amplified characteristics~\cite{kalacheva2025detecting, li2025trust}, while La Morgia et al.~\cite{la2023doge} record more than 900 pump-and-dump events, including DOGE.  
Beyond overt fraud, attention dependence and social coordination further exacerbate fragility: Chen et al.~\cite{chen2025fragility} document strong comovement between TRUMP and MELANIA, and Tandon and Lansiaux~\cite{tandon2021can, lansiaux2022community} link DOGE’s market activity directly to online discourse.  
Overall, memecoins operate under highly concentrated governance, where whale dominance and predatory design lead to asymmetric outcomes, with few extreme winners amid widespread collapse.

\subsection{Feature III: Narrative-based Valuation and Affective Utility}

Finally, memecoins derive value from external narratives and collective affect rather than intrinsic utility.  
Nani~\cite{nani2022doge} shows that DOGE’s cultural roots lowered entry barriers and enabled participation across financial and meme economies.  
As memes circulate, they shift from countercultural vernacular to commodified spectacle~\cite{mitman2025into}, while fandom and celebrity influence become core drivers of price formation~\cite{brichta2023fanning, serada2023happier}.  
Community storytelling on \textit{4chan} and \textit{Reddit} functions as digital folklore that helps investors metabolize volatility~\cite{yogarajah2022hodling}, while social-media coordination creates feedback loops between discourse and price~\cite{tandon2021can, lansiaux2022community}.  

Individual motives blend perceived economic value with entertainment and gambling-like risk preferences~\cite{anton2024moon, philander2023meme}.  
At the data level, Long et al.~\cite{long2025bridging} develop a Pump.fun dataset combining cultural and financial signals, situating memecoins within broader social-finance research~\cite{nobanee2023we}.  
These studies collectively frame memecoins as assets with narrative-based valuation, where attention and affect function as alternative forms of utility.

\smallskip
\textbf{However,} current literature examines volatility, concentration, and sentiment in isolation.  
Despite extensive descriptive work, no unified framework quantifies how these mechanisms jointly shape structural fragility. Our study fills this gap by introducing ME2F, a cross-dimensional framework integrating volatility, whale dominance, and sentiment amplification to evaluate systemic fragility within the memecoin ecosystem. 
\section{The ME2F Measurement Framework} 
\label{sec-me2f}


Building on the features of memecoins identified in the previous section, we design the ME2F to systematically capture fragility within memecoin ecosystems. While volatility, market concentration, and cultural narratives jointly explain the structural instability of memecoins, empirical evaluation requires translating these qualitative mechanisms into quantitative indicators that allow consistent measurement and cross-asset comparison.

Our proposed ME2F delineates these three interrelated scores as the pillars of fragility measurement: the \textit{VDS} captures volatility persistence, extremes, and spillovers; the \textit{WDS} quantifies ownership concentration and inequality among whales; and the \textit{SAS} links sentiment instability to the amplification of market shocks through attention-driven narratives. These scores collectively provide a coherent taxonomy of fragility, and the subsequent subsections define concrete indicators for each dimension.

\subsection{Volatility Dynamics Score (VDS)}

Price volatility plays a central role in shaping memecoin trading patterns, where rapid inflows and withdrawals of capital often accompany sharp price movements. For capturing such behavior systematically, we focus on short-term volatility as the key indicator of instability. 

To quantify intraday price fluctuations, we measure daily volatility using a simplified range-based estimator~\cite{d2022deep}:
\begin{equation}
V_t = \frac{P^{\max}_t - P^{\min}_t}{P^{c}_{t-1}},
\end{equation}
where $P^{\max}_t$ and $P^{\min}_t$ denote the maximum and minimum transaction prices observed on day $t$, respectively, and $P^{c}_{t-1}$ denotes the closing price on day $t-1$. 

To capture these patterns in a unified way, we propose VDS, a quantitative indicator that incorporates adjustments for base-layer spillovers when applicable. This construction enables consistent comparison of volatility-driven fragility across tokens and their underlying ecosystems.

For each token $i$, let $\bar{v}_i$ denote its average daily volatility and $v^{\max}_i$ its maximum daily volatility observed during the sample period. 
We normalize these volatility metrics as:

\begin{align}
    v^{(a)}_i &= \frac{\bar{v}_i}{\max_{j \in \mathcal{T}} \bar{v}_j} \\
    v^{(m)}_i &= \frac{v^{\max}_i}{\max_{j \in \mathcal{T}} v^{\max}_j}
\end{align}
where $\mathcal{T}$ denotes the set of all tokens under consideration. This normalization maps both components to $[0,1]$, making cross-token comparisons scale-free.

To assess both persistent and extreme forms of price instability, we combine normalized average and maximum volatility into a single composite measure:
\begin{equation}
\mathcal{V}_i = \alpha \cdot v^{(a)}_i + (1-\alpha) \cdot v^{(m)}_i
\end{equation}
where $v^{(a)}_i$ reflects persistent fluctuations, $v^{(m)}_i$ captures extreme volatility, and $\alpha \in [0,1]$ balances their contribution. We choose $\alpha=0.5$, which assigns equal importance to persistent and extreme volatility.

To incorporate market scale, let $z_i$ and $c_i$ denote the maximum daily volume and maximum market capitalization of token $i$, respectively. 
We combine them through the harmonic mean, which yields an effective scale governed by the limiting dimension of volume or capitalization:
\begin{equation}
    S_i = \frac{2}{\tfrac{1}{z_i} + \tfrac{1}{c_i}}
\end{equation}
where $S_i$ serves as a market scale factor that is sensitive to both liquidity and capitalization. 

Based on this, we introduce a resilience factor that moderates the fragility of large-scale tokens:
\begin{equation}
    R_i = \frac{1}{1+\gamma S_i}
\end{equation} 
where $\gamma>0$ controls the down-weighting effect. 
We set $\gamma=0.5$ so that large, liquid tokens are moderately down-weighted without overwhelming the volatility signal.

We obtain a scale-adjusted score by proportionally down-weighting the baseline measure with the resilience factor:
\begin{equation}
    \Phi_i = \mathcal{V}_i \cdot R_i
\end{equation}
The score $\Phi_i$ represents volatility-driven fragility net of market scale, where higher values indicate greater vulnerability.

If token $i$ is issued on a base-layer chain, we further adjust for spillover effects by amplifying with both the base chain’s
volatility and scale:
\begin{equation}
    \Phi^{'}_i = 
    \Phi_i \cdot 
    \Bigg(1 + \beta \cdot \frac{v^{(a)}_{\text{base}} + v^{(m)}_{\text{base}}}{2}
     \cdot \ln \!\big(1 + S_{\text{base}}\big)\Bigg)
\end{equation} 
where $v^{(a)}_{\text{base}}$ and $v^{(m)}_{\text{base}}$ are the average and maximum volatility of the base chain, $S_{\text{base}}$ its scale factor, and $\beta>0$ a tuning parameter. We fix $\beta=0.5$ to model a moderate spillover effect. For standalone chains (e.g., DOGE), no adjustment is applied. To compress extreme dispersion, we further rescale the final score through a monotonic transformation:
\begin{equation}
\boxed{
\mathrm{VDS}_i =
  \begin{cases}
    \sqrt{\Phi_i}, & \text{standalone chain} \\
    \sqrt{\Phi^{'}_i}, & \text{base-chain token}
  \end{cases}
}
\end{equation}

\subsection{Whale Dominance Score (WDS)}

Ownership in memecoin ecosystems is typically highly concentrated, with a small set of addresses controlling the majority of circulating supply. Such concentration amplifies manipulation risks and structural inequality. To capture these effects quantitatively, the WDS is designed to measure two complementary dimensions of concentration in memecoin ecosystems. 


First, the aggregate share of the top $n$ addresses reflects their external dominance relative to the overall supply. Second, the distribution among these addresses distinguishes custodial aggregation from genuine whale control. Together, these dimensions highlight structural fragility when dominance and inequality coincide.

For each token $i$, let $o_{k,i}$ denote the ownership share held by the $k$-th largest address, expressed as a fraction of the token’s total supply. The cumulative share of the top $n$ addresses is:
\begin{equation}
C_i = \sum_{k=1}^{n} o_{k,i}
\end{equation}
which captures the external dominance of these addresses relative to the overall supply.

To further distinguish whether this dominance arises from a few large whales or from more balanced holdings, 
we consider the internal distribution among these addresses using the Herfindahl--Hirschman Index (HHI)~\cite{matsumoto2012some}:
\begin{equation}
H_i = \sum_{k=1}^{n} o_{k,i}^2
\end{equation}
which rises as ownership becomes more concentrated, reaching its minimum when holdings are equally distributed across the $n$ addresses and its maximum when a single address controls the entire share.

Given a fixed cumulative share $C_i$, the value of $H_i$ ranges from a minimum to a maximum. We therefore normalize $H_i$ onto the unit interval as:
\begin{equation}
\mathcal{N}_i =
\frac{\tfrac{H_i}{C_i^2} - \tfrac{1}{n}}{1 - \tfrac{1}{n}}
\end{equation}
so that $\mathcal{N}_i$ denotes the degree of internal concentration, independent of scale.

In our analysis, we fix $n=100$, so that $C_i$ reflects the cumulative holdings of the top 100 addresses, while $\mathcal{N}_i$ characterizes the inequality among them. Finally, the concentration score for token $i$ is defined as:
\begin{equation}
\boxed{
\text{WDS}_i = C_i \cdot \mathcal{N}_i
}
\end{equation}

Higher values of WDS signify a greater degree of whale dominance and indicate heightened structural fragility in the ownership configuration of the token.

\subsection{Sentiment Amplification Score (SAS)}

Memecoin valuations are highly sensitive to sentiment interventions and social-media narratives. Endorsements, viral posts, or symbolic events can rapidly amplify attention, triggering sharp inflows followed by abrupt reversals. To quantify sentiment amplification, we use the Fear and Greed Index (FGI) as a sentiment proxy and evaluate token-level responses across benchmark tokens and memecoins

We define the SAS as the combination of two components: the \emph{Instability Index}, which captures baseline sentiment instability; and the \emph{Shock Index}, which measures shock-price transmission strength. We detail each component in turn. Table~\ref{tab:notation_fgi} summarizes the notation of all indicators.  


\begin{table}[!t]
\centering
\caption{FGI-based indicators and corresponding notation}
\label{tab:notation_fgi}
\resizebox{0.9\linewidth}{!}{
\begin{tabular}{ll}
\toprule
\textbf{Symbol} & \textbf{Description} \\
\midrule
$\bar{F}$ & Average FGI value \\
$F_{\max}$ & Maximum FGI value \\
$F_{\min}$ & Minimum FGI value \\
$R_{F}$ & Maximum sentiment range, $F_{\max} - F_{\min}$ \\
$Q_{g}$ & Proportion of samples with FGI of extreme greed \\
$Q_{f}$ & Proportion of samples with FGI of extreme fear \\
$\Delta F_{max}$ & Maximum absolute change in FGI over one day \\
$\Delta P_{max}$ & Maximum absolute change in price over one day \\
\bottomrule
\end{tabular}
}
\end{table}

In \emph{Instability Index ($U$)}, for each token~$i$, let:
\begin{align}
  R_{F,i} &= F_{\max,i}-F_{\min,i} \\
  E_i &= Q_{g,i} + Q_{f,i} \\
  \bar F_i &= \tfrac{1}{T}\sum_{t=1}^{T}F_{i,t}
\end{align}
where $F_{i,t}$ denotes the coin-specific FGI at time~$t$, 
and $T$ is the total number of observations in the sample period. We then normalize each quantity by its cross-sectional maximum and take the arithmetic mean:
\begin{equation}
U_i
  = \tfrac13\!\left(
      \frac{R_{F,i}}{\max_j R_{F,j}}
      +\frac{E_i}{\max_j E_j}
      +\frac{|\bar F_i-50|}{\max_j |\bar F_j-50|}
    \right)
\end{equation}
Tokens with large sentiment ranges, frequent extreme states, or a persistent bias away from the neutral mid-point receive higher $U_i$ scores.

In \emph{Shock Index} ($K$), we define $\Delta F^{\max}_i$ as the largest single-day absolute change in FGI and $\Delta P^{\max}_i$ the corresponding absolute price return. Their joint effect is:
\begin{equation}
K_i = 
    \frac{\Delta F^{\max}_i}{\max_j \Delta F^{\max}_j}
    \cdot
    \frac{\Delta P^{\max}_i}{\max_j \Delta P^{\max}_j}
\end{equation}
which attains high values only when both the sentiment shock and the corresponding price response are simultaneously pronounced.

Based on the above analysis, we formalize SAS as:
\begin{equation}
\boxed{\text{SAS}_i =
    U_i \cdot (K_i)^{\delta}}
\end{equation}
where $U_i$ preserves the role of baseline sentiment instability, while the parameter $\delta$ governs how strongly shock-price transmission is amplified. In our experiments, we set $\delta = 1.5$, which emphasizes that abrupt shocks are treated as more influential than gradual sentiment drifts in shaping price fragility.

\section{Evaluation Results}
\label{sec-me2f}

Building on the ME2F framework introduced previously, we apply the three indicators to a representative sample of major memecoins to evaluate fragility patterns in practice.
The analysis first examines descriptive characteristics of market volatility, ownership concentration, and sentiment dynamics across seven representative tokens, including DOGE, SHIB, TRUMP, MELANIA, FLOKI, PEPE, and LIBRA, with two benchmark tokens, ETH and SOL. We then operationalize the ME2F framework to quantify fragility along these three dimensions.

The analysis relies on data from multiple sources, including \emph{CoinMarketCap}~\cite{coinmarketcap}, \emph{CoinGecko}~\cite{coingecko}, \emph{Etherscan}~\cite{etherscan}, and \emph{CoinCarp}~\cite{coincarp}, spanning time windows across 2023--2025. We next apply the ME2F framework to assess fragility across its three dimensions.

\begin{table*}[!]
    \centering
    \caption{Volatility and market metrics summary}
    \resizebox{0.85\linewidth}{!}{
    \begin{threeparttable}
    \begin{tabular}{l
                S[table-format=3.2,round-mode=places,round-precision=2]
                S[table-format=3.2,round-mode=places,round-precision=2]
                S[table-format=4.2,round-mode=places,round-precision=2]
                S[table-format=5.2,round-mode=places,round-precision=2]}
        \toprule
        \textbf{Token} & {\textbf{Avg. Volatility (\%)}} & {\textbf{Max. Volatility (\%)}} & {\textbf{Max. Daily Volume (USD billions)}} & {\textbf{Max. Market Cap (USD billions)}} \\
        \midrule
        DOGE    &  6.27 &  41.19  & 399.36  & 687.49  \\
        SHIB    &  6.35 &  63.36  & 160.15  & 211.52  \\
        TRUMP   & 15.26 &  62.25  & 390.64  &  91.01  \\
        MELANIA & 11.94 &  45.24  &  56.30  &   9.15  \\
        FLOKI   &  9.49 &  80.94  &  24.55  &  30.60  \\
        PEPE    & 12.92 & 301.77  & 156.94  & 111.08  \\
        LIBRA   & 12.63 & 126.83  &  16.68  &   1.19  \\
        \midrule
        ETH     &  4.32 &  26.50  & 924.54  & 4884.02 \\
        SOL     &  6.70 &  26.70  & 331.73  & 1274.19 \\
        \bottomrule
    \end{tabular}
    \end{threeparttable}
    }
    \label{tab:volatility}
\end{table*}

\subsection{Volatility Dynamics}

\begin{figure}[h]
    \centering
    \subfloat[Benchmark tokens]{%
        \includegraphics[width=\columnwidth]{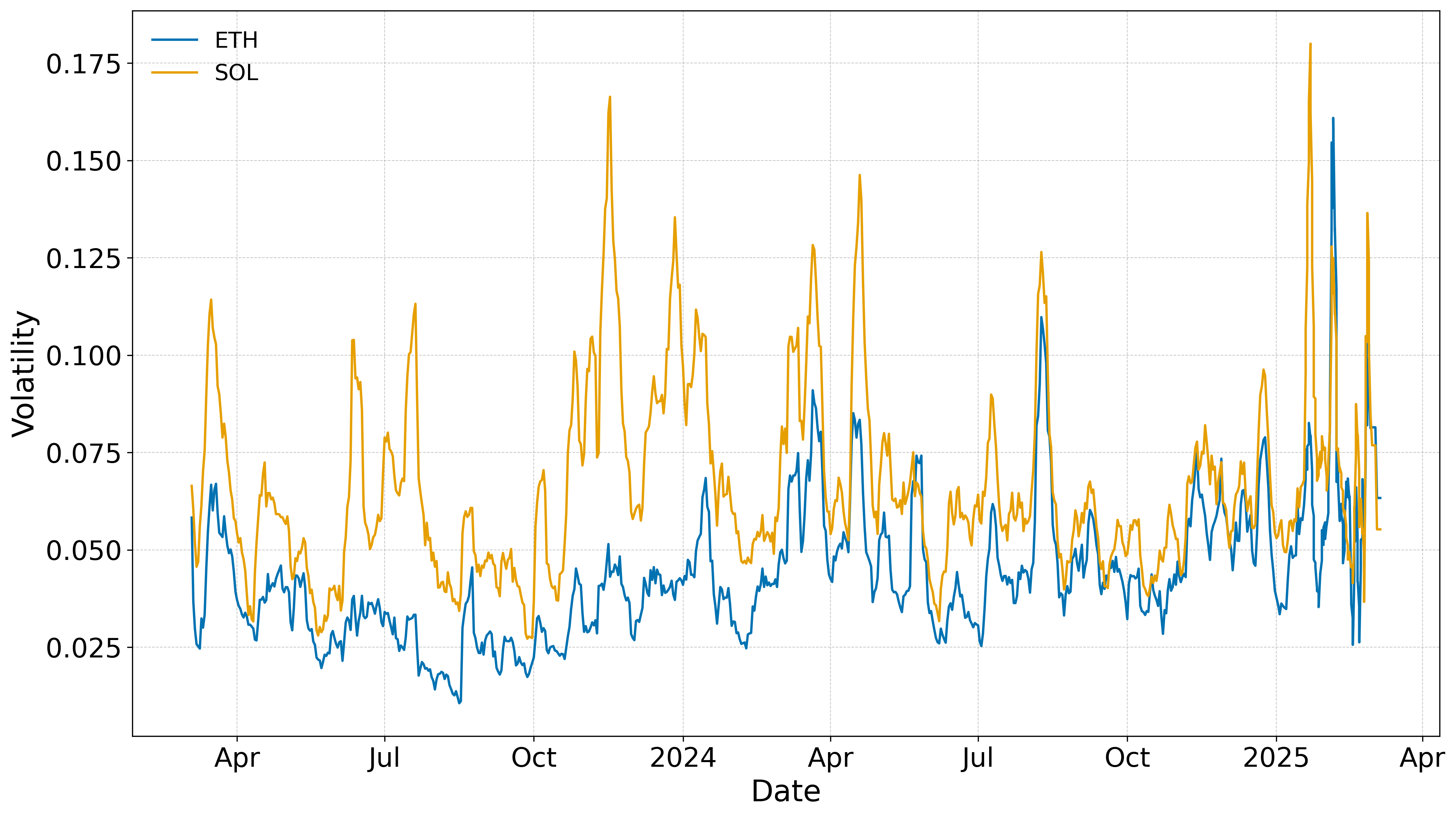}}\\[0.5em]
    \subfloat[Memecoins]{%
        \includegraphics[width=\columnwidth]{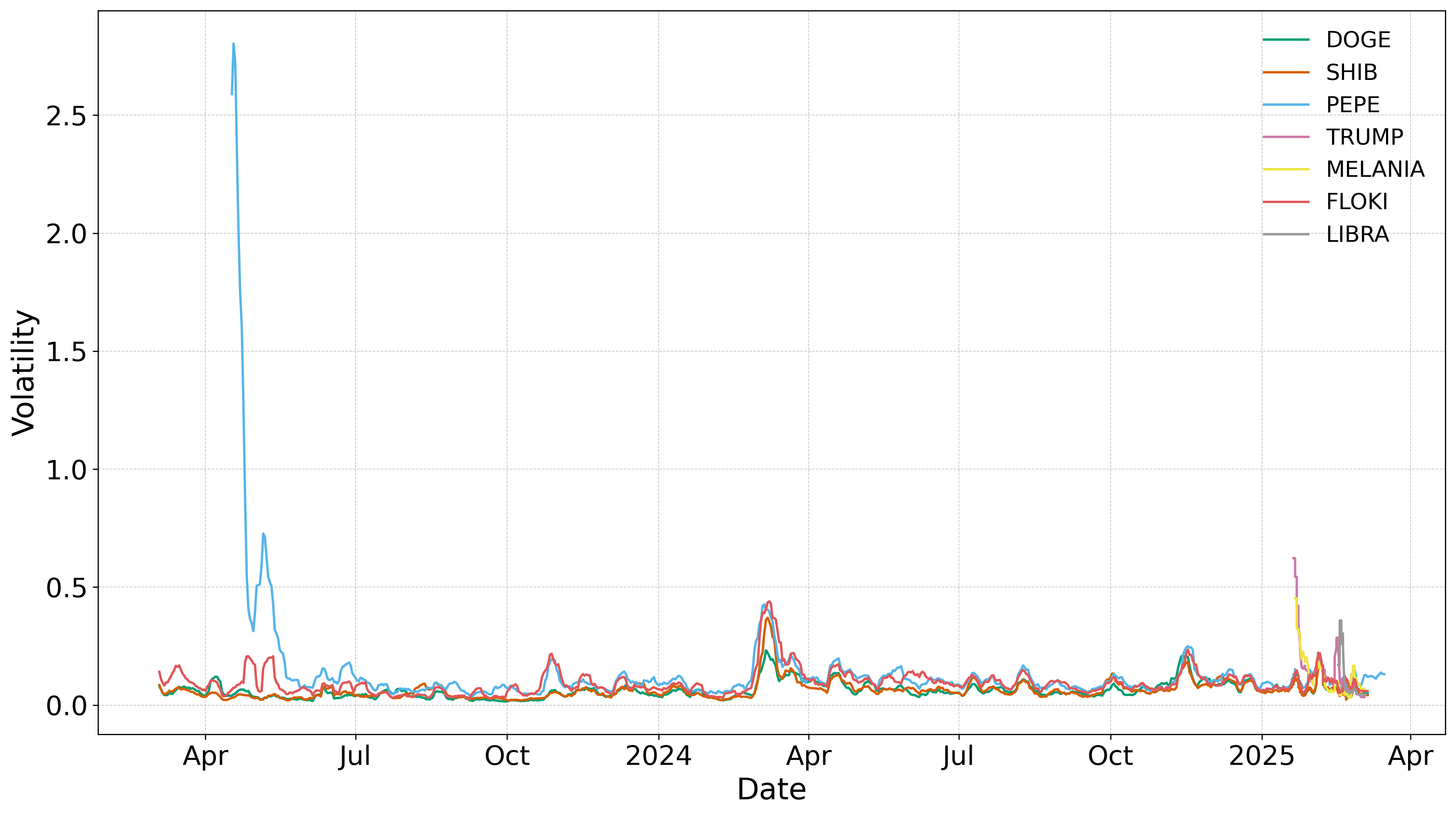}}
    \caption{Temporal dynamics of volatility across various cryptocurrencies.}
    \label{fig:volatility_compact}
\end{figure}

\noindent\textbf{Empirical Analysis of Volatility.} We examine volatility dynamics by linking fluctuations in market value to short-cycle capital flows. Based on this estimator, we compute daily volatility for all tokens in the sample. Table~\ref{tab:volatility} summarizes the results together with market-scale indicators, reporting average and maximum volatility as well as peak daily trading volume and market capitalization. 
Fig.~\ref{fig:volatility_compact} complements this view by showing the temporal evolution of volatility that highlights both synchronized bursts of instability and subgroup-specific patterns that unfold over time.

Empirical evidence across representative memecoins shows that high-volatility episodes are tightly coupled with bursts of trading activity and short-cycle inflows. For instance, DOGE reached a volatility of 29\% on November 2024, when daily volume hit a record of \$39.9B and market capitalization soon climbed to \$68.7B in the next month. SHIB in March 2024 displayed volatility of 63.4\% with volume of \$9.0B, while PEPE peaked at an extreme 301.8\% volatility in April 2023, coinciding with a market-cap surge above \$11B. 

Other memecoins exhibit the same dynamics, though with varying intensity. TRUMP recorded volatility of 62.3\% and daily trading volume of \$30.1B on January 2025, while MELANIA reached a peak volume of \$5.6B and market capitalization of \$0.9B within the same month, showing rapid short-term inflows. FLOKI surged by more than 73.8\% in a single day in March 2024, with trading volume exceeding \$2.3B and sustained activity in the following days. LIBRA displayed a smaller absolute scale but similar behavior, reaching volatility above 120\% and a market-cap peak over \$0.1B in February 2025. 

In addition, memecoin volatility can spill over into the trading dynamics of base-layer tokens, amplifying fluctuations in on-chain activity. For instance, SHIB's peak volatility of 63.4\% had a limited contemporaneous effect on ETH, with ETH volatility at only 5.6\% and volume of \$26.8B on that day. Yet during SHIB's trading-volume peak \$16.0B on 5 March 2024, ETH volatility rose sharply to 16.6\% and volume nearly doubled to \$47.7B, suggesting that capital flows into SHIB may have contributed to swings in ETH activity. Similarly, SOL, as the base chain for TRUMP and other tokens, displayed heightened co-movement. On 19 January 2025, TRUMP's daily volume peaked at \$39.1B, coinciding with a surge in SOL's volume to \$33.2B and volatility to 20.9\%.

Overall, these cases show that memecoin markets are marked by extreme and persistent volatility, where short-lived inflows and speculative bursts drive sharp swings in price and liquidity. Unlike mainstream cryptocurrencies, whose fluctuations are often tied to macroeconomic or technological factors, memecoin volatility reflects structural dependence on sentiment-driven capital and shallow market depth. Crucially, these shocks do not remain confined to individual tokens but frequently spill over into base-layer tokens, amplifying fluctuations in their underlying ecosystems. 

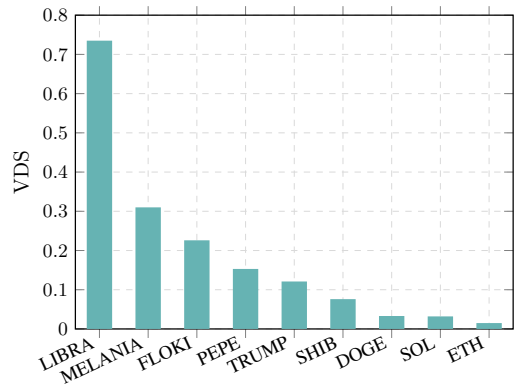
\begin{figure}[!t]
\centering
\begin{tikzpicture}[scale=0.80]
\begin{axis}[
    ybar,
    width=\linewidth,
    height=6.8cm,
    bar width=12pt,
    ymin=0, ymax=0.8,
    ytick={0,0.1,0.2,0.3,0.4,0.5,0.6,0.7,0.8},
    yticklabel style={font=\small},
    ylabel={$\mathrm{VDS}$},
    symbolic x coords={LIBRA,MELANIA,FLOKI,PEPE,TRUMP,SHIB,DOGE,SOL,ETH},
    xtick=data,
    xticklabel style={rotate=25, anchor=east, font=\small},
    enlarge x limits={abs=0.4cm},  
    grid=major,
    grid style={dashed,gray!30},
    axis line style={semithick},
]
  \addplot+[draw=none, fill=teal!60] coordinates {
    (LIBRA,0.735)
    (MELANIA,0.310)
    (FLOKI,0.226)
    (PEPE,0.153)
    (TRUMP,0.121)
    (SHIB,0.0760)
    (DOGE,0.0330)
    (SOL,0.0320)
    (ETH,0.0150)
  };
\end{axis}
\end{tikzpicture}
\vspace{-1em}
\caption{$\mathrm{VDS}$ across cryptocurrencies.}
\label{fig:vfs}
\end{figure}

\smallskip
\noindent\textbf{ME2F-based Volatility Fragility.}  
Fig.~\ref{fig:vfs} presents $\mathrm{VDS}$ values for benchmark tokens and representative memecoins. 
The scores span several orders of magnitude, with ETH, SOL, and DOGE concentrated at the lower end, while Ethereum-hosted tokens SHIB, PEPE, FLOKI show higher fragility and Solana-hosted tokens TRUMP, MELANIA, LIBRA climb further. 
This gradation reveals a systematic divide: deep and liquid base-layer tokens display resilience, established memecoins occupy an intermediate range, and newer or thematically driven tokens rise to the top, with LIBRA marking the extreme.

\begin{figure}[b]
    \centering
    \includegraphics[width=0.9\columnwidth]{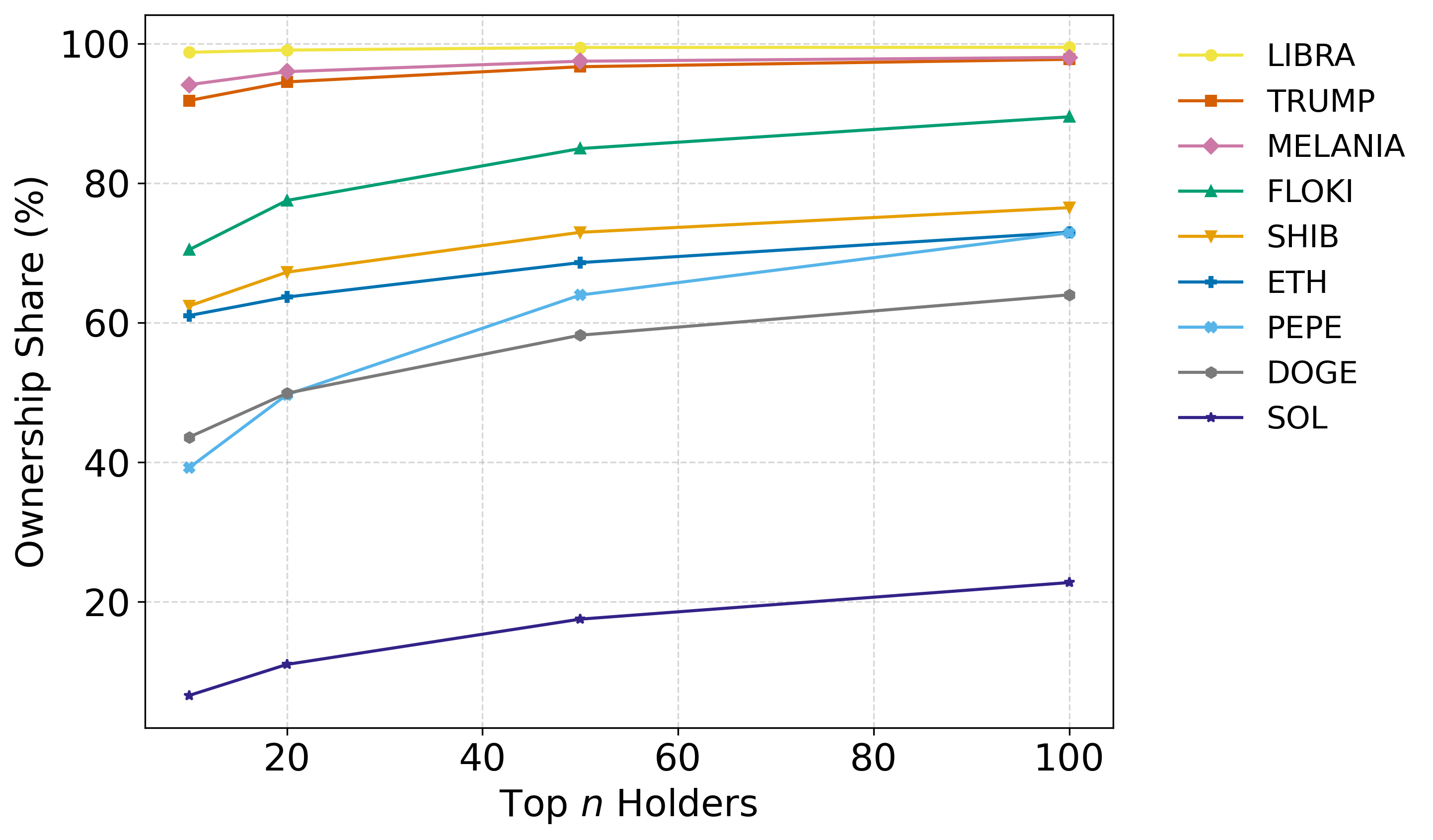}
    \caption{Ownership concentration across various cryptocurrencies.}
    \label{fig:holder_distribution}
\end{figure}

Several structural factors explain this dispersion. 
Benchmark tokens ETH and SOL moderate volatility with large capitalization and liquidity buffers, leading to low fragility. 
DOGE remains relatively resilient due to its broad adoption and deeper markets compared with newer tokens. 
In contrast, SHIB, PEPE, and FLOKI combine frequent speculative spikes with limited depth, yielding elevated scores. 
Politically themed tokens such as TRUMP and MELANIA add episodic attention shocks, while LIBRA illustrates the extreme case where a very small market base cannot absorb volatility, resulting in the highest fragility observed.  

\subsection{Whale Dominance}


\noindent\textbf{Empirical Analysis of Concentration.} As shown in Fig.~\ref{fig:holder_distribution}, mainstream cryptocurrencies provide a useful baseline for interpreting concentration patterns in memecoin ecosystems. Ethereum’s top 100 addresses control about 73\% of supply, a figure largely explained by exchange custody and staking pools that aggregate the assets of many users. Solana, by contrast, exhibits only 23\% concentration, reflecting its smaller scale and broader token distribution, while Dogecoin sits in between at 64\%. These baseline figures illustrate that concentration levels vary significantly even among established cryptocurrencies, depending on institutional arrangements and ecosystem structure.

Memecoins, however, display consistently sharper concentration gradients. FLOKI and SHIB stand out, with their top 100 holders retaining nearly 90\% and 76\% of supply, while PEPE, despite its retail-driven image, still shows 73\% concentration at the top 100 addresses. At the extreme, Solana-based tokens such as TRUMP, MELANIA, and LIBRA reach unprecedented levels, with 98\% or more of total supply locked in the hands of a few wallets. These distributions highlight key structural features of memecoin ecosystems: ownership is dominated by a small number of large addresses, and governance capacity is effectively concentrated in the hands of whales. 

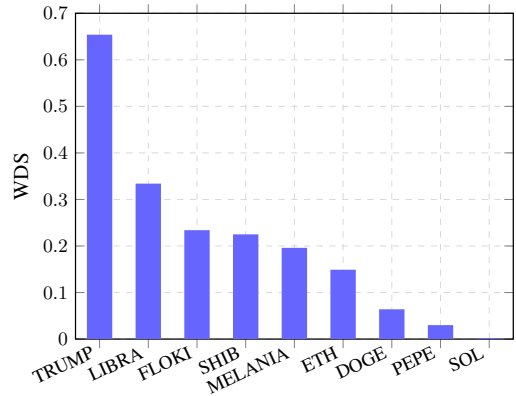
\begin{figure}[!t]
\centering
\begin{tikzpicture}[scale=0.80]
\begin{axis}[
    ybar,
    width=\linewidth,
    height=7cm,
    bar width=12pt,
    ymin=0, ymax=0.7,
    ytick={0,0.1,0.2,0.3,0.4,0.5,0.6,0.7},
    yticklabel style={font=\small},
    ylabel={WDS},
    symbolic x coords={TRUMP,LIBRA,FLOKI,SHIB,MELANIA,ETH,DOGE,PEPE,SOL},
    xtick=data,
    xticklabel style={rotate=25, anchor=east, font=\small},
    enlarge x limits={abs=0.4cm},  
    grid=major,
    grid style={dashed,gray!30},
    axis line style={semithick},
]
  \addplot+[draw=none, fill=blue!60] coordinates {
    (TRUMP,0.654)
    (LIBRA,0.334)
    (FLOKI,0.234)
    (SHIB,0.225)
    (MELANIA,0.196)
    (ETH,0.149)
    (DOGE,0.064)
    (PEPE,0.030)
    (SOL,0.002)
  };
\end{axis}
\end{tikzpicture}
\vspace{-1em}
\caption{WDS across various cryptocurrencies.}
\label{fig:dominance_score}
\end{figure}

\smallskip \noindent\textbf{ME2F-based Concentration Fragility.} Fig.~\ref{fig:dominance_score} presents $\mathrm{WDS}$ values for benchmark tokens and representative memecoins. The scores cluster into distinct tiers: TRUMP dominates with the highest value, followed by LIBRA, while FLOKI, SHIB, and MELANIA occupy the intermediate range. In contrast, ETH, DOGE, PEPE, and especially SOL display substantially lower scores, indicating comparatively diffuse ownership structures. This separation reveals a phenomena: Solana-based political tokens exhibit the sharpest whale dominance, mid-tier memecoins retain significant concentration despite retail narratives, and benchmark cryptocurrencies remain the least concentrated. 

Several structural factors explain this dispersion. High values for TRUMP and LIBRA reflect extremely concentrated supply bases, where a few large wallets exert disproportionate control. FLOKI and SHIB, although widely traded, still exhibit notable concentration among leading holders, sustaining medium fragility. By contrast, ETH and SOL benefit from broad market participation and institutional liquidity, which dilute the relative power of top holders and provide greater resilience against abrupt shifts.
DOGE’s long-standing adoption base similarly limits extreme dominance, as its supply has become more diffused over time despite the asset’s memecoin origin and periodic speculative surges.

\subsection{Sentiment Amplification}


\noindent\textbf{Empirical Analysis of Sentiment.} To implement this dimension, FGI is applied to quantify sentiment amplification across benchmark tokens and representative memecoins. The FGI is partitioned into five ranges:
$0$--$19$ (\emph{Extreme Fear}), 
$20$--$39$ (\emph{Fear}), 
$40$--$59$ (\emph{Neutral}), 
$60$--$79$ (\emph{Greed}), 
and $80$--$100$ (\emph{Extreme Greed}). 

\begin{figure}[!t]
    \centering
    \subfloat[Benchmark tokens]{%
        \includegraphics[width=\columnwidth]{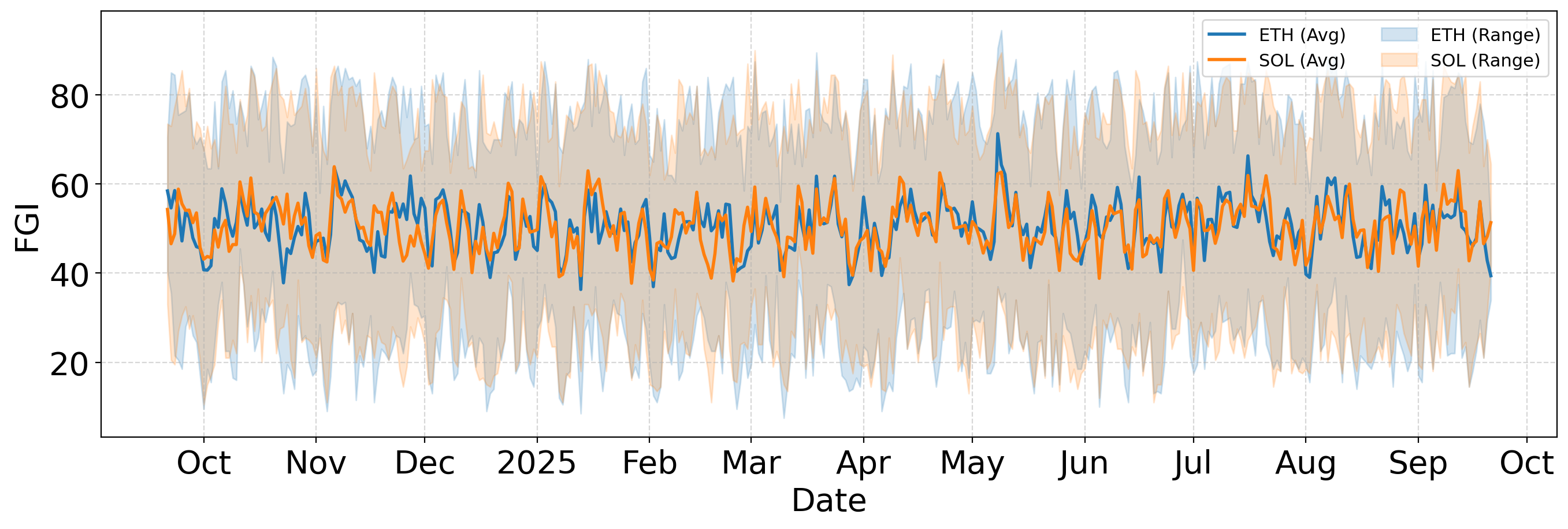}}\\[0.5em]
    \subfloat[Memecoins]{%
        \includegraphics[width=\columnwidth]{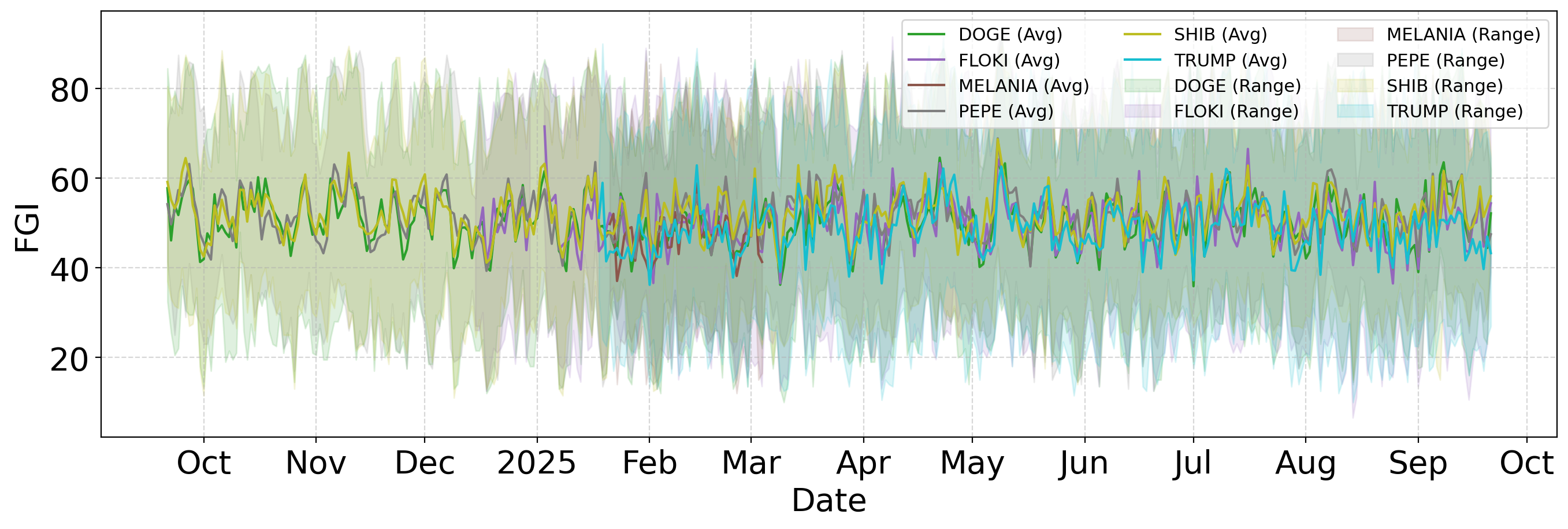}}
    \caption{Temporal patterns of FGI across various cryptocurrencies.}
    \label{fig:fgi_compact}
\end{figure}

For each token, we derive a set of indicators from FGI to characterize baseline sentiment, short-term fluctuations, and extreme states. Fig.~\ref{fig:fgi_compact} shows that FGI for ETH and SOL fluctuates within a relatively narrow band. Daily averages remain close to the neutral zone, and intraday min–max ranges are comparatively contained. The two tokens also exhibit broadly similar temporal patterns, which appear to reflect shifts in broader market sentiment and liquidity conditions. Overall, their sentiment dynamics suggest a more stable profile, though variations are still present.

By comparison, memecoins display larger swings in the FGI, with wider intraday ranges. The daily averages move more sharply across time, and extreme values frequently deviate from the neutral zone. Certain tokens, such as PEPE and SHIB, experience distinct short-lived peaks, while others, such as TRUMP and MELANIA, occasionally dip into extended low-sentiment phases. These descriptive patterns highlight the stronger fluctuations in memecoin sentiment relative to benchmark tokens. 

Complementing Fig.~\ref{fig:fgi_compact}, Table~\ref{tab:fgi_summary} highlights two main contrasts.  Sentiment ranges are wide across all tokens, with ETH showing the largest span \(R_F=87.0\) and MELANIA the narrowest \(73.5\). However, ETH and SOL display more balanced greed-fear occupancy, whereas SHIB and PEPE are strongly greed-skewed. This indicates that benchmarks, despite occasionally wide sentiment ranges, sustain more symmetric profiles overall.  

\begin{table}[!t]
\centering
\caption{FGI-based metric summary}
\label{tab:fgi_summary}
\resizebox{\linewidth}{!}{%
\begin{tabular}{lcccccccc}
\toprule
Token & $\bar{F}$ & $F_{\max}$ & $F_{\min}$ & $R_F$ & $Q_{g}$ & $Q_{f}$ & $\Delta F_{max}$ & $\Delta P_{max}$ \\
\midrule
DOGE    & 50.41 & 92.00 & 10.00 & 82.00 & 1.23\% & 0.71\% & 52.50 & 8.07\% \\
SHIB    & 51.95 & 89.50 & 11.00 & 78.50 & 0.90\% & 0.21\% & 55.00 & 11.27\% \\
TRUMP   & 48.76 & 91.50 & 10.00 & 81.50 & 0.84\% & 0.89\% & 50.50 & 22.40\% \\
MELANIA & 46.76 & 85.50 & 12.00 & 73.50 & 0.62\% & 0.75\% & 40.50 & 14.95\% \\
FLOKI   & 50.29 & 91.50 &  6.50 & 85.00 & 0.92\% & 0.80\% & 46.50 & 10.07\% \\
PEPE    & 51.74 & 93.00 & 12.50 & 80.50 & 1.04\% & 0.15\% & 45.50 & 14.80\% \\
\midrule
ETH     & 50.60 & 94.50 &  7.50 & 87.00 & 1.19\% & 1.02\% & 53.50 & 10.12\% \\
SOL     & 50.45 & 90.00 & 10.50 & 79.50 & 0.93\% & 0.65\% & 55.50 & 8.80\% \\
\bottomrule
\end{tabular}%
}
\end{table}

The price impact of sentiment shocks is where differences become sharper. TRUMP records the largest response \(\Delta P_{\max}=22.4\%\), with MELANIA and PEPE also elevated, 
while ETH and SOL show smaller adjustments around 10\%. 
DOGE is similarly low at 8.07\%, reflecting its deeper liquidity and longer adoption history.  
Thus, while benchmarks can experience large sentiment swings, memecoins are generally more likely to amplify them into outsized price effects.  

\begin{figure}[h]
\centering
\begin{tikzpicture}[scale=0.80]
\begin{axis}[
    ybar,
    width=\linewidth,
    height=7cm,
    bar width=12pt,
    ymin=0, ymax=0.7,
    ytick={0,0.1,0.2,0.3,0.4,0.5,0.6,0.7},
    yticklabel style={font=\small},
    ylabel={SAS},
    symbolic x coords={TRUMP,MELANIA,PEPE,SHIB,ETH,SOL,FLOKI,DOGE},
    xtick=data,
    xticklabel style={rotate=25, anchor=east, font=\small},
    enlarge x limits={abs=0.4cm},
    grid=major,
    grid style={dashed,gray!30},
    axis line style={semithick},
]
  \addplot+[draw=none, fill=orange!70] coordinates {
(TRUMP,0.608) 
(MELANIA,0.279)
(PEPE,0.266)
(SHIB,0.235)
(ETH,0.209)
(SOL,0.145)
(FLOKI,0.142)
(DOGE,0.129)
};
\end{axis}
\end{tikzpicture}
\vspace{-1em}
\caption{SAS across various cryptocurrencies.}
\label{fig:sas_score}
\end{figure}
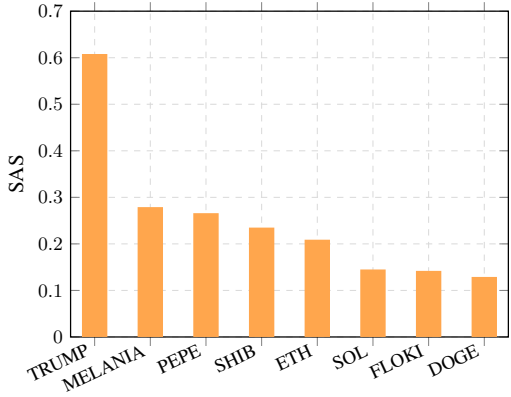

\smallskip
\noindent\textbf{ME2F-based Sentiment Fragility.}
Fig.~\ref{fig:sas_score} reports $\mathrm{SAS}$ values for benchmark tokens and representative memecoins.
The distribution shows a steep gradient: TRUMP stands out with the highest score, far exceeding all others, while MELANIA, PEPE, and SHIB form a secondary cluster in the mid-range.
By contrast, ETH, SOL, FLOKI, and DOGE register substantially lower values, indicating markedly stronger resilience.
This separation highlights a systematic divide: politically themed tokens, especially TRUMP, exhibit pronounced sensitivity to short-lived shocks, mid-tier memecoins show moderate instability, and benchmark tokens anchor the lower end of the scale.

Several structural factors explain this dispersion.
The extreme value for TRUMP reflects its narrow liquidity base and high exposure to episodic attention shocks, making it highly susceptible to transient swings.
MELANIA, PEPE, and SHIB sustain intermediate fragility, combining speculative trading surges with limited market depth.
In contrast, ETH and SOL mitigate shock amplification through large capitalization and institutional liquidity, which cushion external disturbances.
DOGE, despite its memecoin origin, maintains resilience owing to its long adoption history and broad distribution, while FLOKI’s somewhat wider retail participation prevents sharp concentration effects but leaves it more fragile than benchmark tokens.

\section{Discussion}
\label{sec-discussion}


\subsection{Cross-metric Comparison and Layered Patterns} 

As shown in Table~\ref{tab:fragility_all}, the three fragility indicators reveal a stratified landscape. TRUMP stands out with uniformly extreme values across VDS, WDS, and SAS, making it the most fragile token in the sample. LIBRA and MELANIA follow in the upper tier with elevated VDS and WDS, though LIBRA lacks SAS data. At the opposite end, DOGE, SOL, and ETH occupy the low-fragility tier, combining small values across all metrics and confirming their systemic resilience.  

Between these extremes lie SHIB, PEPE, and FLOKI, which present mixed profiles. SHIB shows moderately elevated volatility and amplification, FLOKI exhibits high volatility but weak amplification, and PEPE combines strong volatility and amplification with relatively low whale concentration. Overall, political tokens on Solana (TRUMP, MELANIA, LIBRA) cluster at the fragile extreme, base-layer assets (ETH, SOL, DOGE) define the stable tier, and community-driven tokens (SHIB, PEPE, FLOKI) occupy the middle ground.

\smallskip
\subsection{Interactions and Variations across Dimensions}

The three indicators do not always move in tandem, clarifying the sources of fragility. High volatility often aligns with strong sentiment amplification; TRUMP, SHIB, and PEPE exemplify this, suggesting that tokens prone to sharp price swings also amplify collective sentiment shocks. However, ownership concentration frequently diverges from these dynamics, revealing distinct structural risks.  

PEPE illustrates this asymmetry most clearly, combining elevated volatility and amplification with dispersed ownership. Conversely, LIBRA exhibits strong volatility and concentration but lacks amplification data, showing that governance risks can persist independently. FLOKI provides another contrast, with relatively high volatility but weak amplification. These divergences confirm that fragility is multi-faceted, highlighting the need for a multi-indicator approach to assessment.

\begin{table}[b]
\centering
\caption{Fragility measurement summary across tokens}
\label{tab:fragility_all}
\small
\resizebox{0.85\linewidth}{!}{
\begin{tabular}{lccc}
\toprule
\textbf{Token} & \textbf{VDS} & \textbf{WDS} & \textbf{SAS} \\
\midrule
ETH     & \heatTeal{0}{\(1.50\times10^{-2}\)} & \heatBlue{23}{\(1.49\times10^{-1}\)} & \heatOrange{17}{\(2.09\times10^{-1}\)} \\
SOL     & \heatTeal{4}{\(3.20\times10^{-2}\)} & \heatBlue{0}{\(2.00\times10^{-3}\)}  & \heatOrange{3}{\(1.45\times10^{-1}\)} \\
DOGE    & \heatTeal{5}{\(3.30\times10^{-2}\)} & \heatBlue{10}{\(6.40\times10^{-2}\)} & \heatOrange{0}{\(1.29\times10^{-1}\)} \\
SHIB    & \heatTeal{11}{\(7.60\times10^{-2}\)} & \heatBlue{34}{\(2.25\times10^{-1}\)} & \heatOrange{22}{\(2.35\times10^{-1}\)} \\
TRUMP   & \heatTeal{18}{\(1.21\times10^{-1}\)} & \heatBlue{100}{\(6.54\times10^{-1}\)}& \heatOrange{100}{\(6.08\times10^{-1}\)} \\
PEPE    & \heatTeal{24}{\(1.53\times10^{-1}\)} & \heatBlue{4}{\(3.00\times10^{-2}\)}  & \heatOrange{29}{\(2.66\times10^{-1}\)} \\
FLOKI   & \heatTeal{33}{\(2.26\times10^{-1}\)} & \heatBlue{36}{\(2.34\times10^{-1}\)} & \heatOrange{3}{\(1.42\times10^{-1}\)} \\
MELANIA & \heatTeal{45}{\(3.10\times10^{-1}\)} & \heatBlue{30}{\(1.96\times10^{-1}\)} & \heatOrange{31}{\(2.79\times10^{-1}\)} \\
LIBRA   & \heatTeal{100}{\(7.35\times10^{-1}\)} & \heatBlue{51}{\(3.34\times10^{-1}\)} & \multicolumn{1}{c}{—} \\
\bottomrule
\addlinespace[3pt]
\end{tabular}
}
\parbox{.80\linewidth}{\footnotesize
Notes: Darker shading indicates higher fragility.}
\end{table}

\subsection{Application as Early-Warning Signals}
ME2F can function like a risk alarm system. We track each score, i.e., VDS, WDS, and SAS, over a rolling window and raise a flag when a score becomes unusually high relative to its own recent history, e.g., when it enters the top 10\% of its trailing range. We also flag it as trouble when two scores spike close together within $x$ days.

Tokens driven primarily by attention, such as TRUMP and MELANIA, tend to breach these thresholds far more frequently, whereas deep and liquid benchmarks like ETH and SOL rarely do. This contrast aligns with the intuition underlying ME2F’s design.

For practical application, we propose three action buckets. If \emph{VDS and SAS are both high}, tighten risk: reduce exposure, slow position growth, and watch order books more closely. If \emph{WDS is high}, focus on governance and whale risk: track top holders, monitor unlock calendars, and prepare for block sales. If \emph{all scores are low}, keep standard monitoring.

\smallskip
\subsection{Implications for Ecosystems and Governance}

The findings carry implications for governance and regulation. On the governance side, extreme whale concentration exposes ecosystems to capture, in tension with retail-driven narratives. On the regulatory side, the clustering of high volatility and amplification highlights the acute sensitivity of memecoin markets to external narratives, particularly those amplified on social media.  

From a market perspective, fragility profiles fall into distinct categories. Political tokens (TRUMP, MELANIA, LIBRA) carry systemic vulnerabilities across multiple dimensions. Community-driven tokens (SHIB, PEPE, FLOKI) show medium fragility shaped by episodic speculation and shallow liquidity. The older token (DOGE) demonstrates relative resilience, illustrating how longevity and broad participation can partially insulate a memecoin from extreme fragility.

\subsection{ME2F Limitations}

ME2F offers clear signals but faces limits in data and scope. The current implementation leans on aggregated market and sentiment data and skips structural layers like transaction topology and cross-chain migration. Short time windows curb long-run analysis, and separating on-chain from off-chain signals hides how microstructure interacts with narrative shocks.

Next, link on-chain structure to off-chain behavior, extend the time horizon to cover multiple cycles, and fuse transaction networks and cross-chain flows with external attention metrics. These additions would give a holistic view of memecoin ecosystems and let ME2F trace how structural stress and narrative-driven behavior reinforce each other.

\section{Conclusion}
This paper proposes the first Memecoin Ecosystem Fragility Framework (ME2F) to capture memecoin fragility through volatility dynamics, whale dominance, and sentiment amplification. Empirical results show a clear stratification: political tokens are most fragile, established memecoins occupy the middle ground, and benchmark tokens remain resilient.

\bibliographystyle{IEEEtran}
\bibliography{bib}

\begin{thebibliography}{10}
\providecommand{\url}[1]{#1}
\csname url@samestyle\endcsname
\providecommand{\newblock}{\relax}
\providecommand{\bibinfo}[2]{#2}
\providecommand{\BIBentrySTDinterwordspacing}{\spaceskip=0pt\relax}
\providecommand{\BIBentryALTinterwordstretchfactor}{4}
\providecommand{\BIBentryALTinterwordspacing}{\spaceskip=\fontdimen2\font plus
\BIBentryALTinterwordstretchfactor\fontdimen3\font minus \fontdimen4\font\relax}
\providecommand{\BIBforeignlanguage}[2]{{%
\expandafter\ifx\csname l@#1\endcsname\relax
\typeout{** WARNING: IEEEtran.bst: No hyphenation pattern has been}%
\typeout{** loaded for the language `#1'. Using the pattern for}%
\typeout{** the default language instead.}%
\else
\language=\csname l@#1\endcsname
\fi
#2}}
\providecommand{\BIBdecl}{\relax}
\BIBdecl

\bibitem{shifman2013memes}
L.~Shifman, \emph{Memes in digital culture}.\hskip 1em plus 0.5em minus 0.4em\relax MIT press, 2013.

\bibitem{conlon2025memecoin}
T.~Conlon and S.~Corbet, ``Memecoin contagion: Irrationality, illicit behaviour, and cryptocurrency risk,'' \emph{Finance Research Letters}, p. 108264, 2025.

\bibitem{kalacheva2025detecting}
A.~Kalacheva, P.~Kuznetsov, I.~Vodolazov, and Y.~Yanovich, ``Detecting rug pulls in decentralized exchanges: The rise of meme coins,'' \emph{Blockchain: Research and Applications}, p. 100336, 2025.

\bibitem{coingecko}
\BIBentryALTinterwordspacing
{CoinGecko}. (2025) Coingecko. Accessed: Sep. 21, 2025. [Online]. Available: \url{https://www.coingecko.com}
\BIBentrySTDinterwordspacing

\bibitem{taffler2024narrative}
R.~J. Taffler, V.~Agarwal, and M.~Obring, ``Narrative emotions and market crises,'' \emph{Journal of Behavioral Finance}, pp. 1--21, 2024.

\bibitem{nani2022doge}
A.~Nani, ``The doge worth 88 billion dollars: A case study of dogecoin,'' \emph{Convergence}, vol.~28, no.~6, pp. 1719--1736, 2022.

\bibitem{chen2025fragility}
M.~Chen, S.~Wang, and Y.~Wei, ``Fragility of trump and melania coins,'' \emph{Finance Research Letters}, p. 108134, 2025.

\bibitem{zhang2021popular}
S.~Zhang and G.~Mani, ``Popular cryptoassets (bitcoin, ethereum, and dogecoin), gold, and their relationships: Volatility and correlation modeling,'' \emph{Data Science and Management}, vol.~4, pp. 30--39, 2021.

\bibitem{aloosh2022bubbles}
A.~Aloosh, S.~Ouzan, and S.~J.~H. Shahzad, ``Bubbles across meme stocks and cryptocurrencies,'' \emph{Finance Research Letters}, vol.~49, p. 103155, 2022.

\bibitem{yousaf2023connectedness}
I.~Yousaf, L.~Pham, and J.~W. Goodell, ``The connectedness between meme tokens, meme stocks, and other asset classes: Evidence from a quantile connectedness approach,'' \emph{Journal of International Financial Markets, Institutions and Money}, vol.~82, p. 101694, 2023.

\bibitem{chopra2024does}
M.~Chopra, C.~Mehta, P.~Lal, and A.~Srivastava, ``Does the big boss of coins—bitcoin—protect a portfolio of new-generation cryptos? evidence from memecoins, stablecoins, nfts and defi,'' \emph{China Finance Review International}, vol.~14, no.~3, pp. 480--521, 2024.

\bibitem{li2022will}
C.~Li and H.~Yang, ``Will memecoins’ surge trigger a crypto crash? evidence from the connectedness between leading cryptocurrencies and memecoins,'' \emph{Finance Research Letters}, vol.~50, p. 103191, 2022.

\bibitem{li2023can}
Y.~Li, B.~Lucey, and A.~Urquhart, ``Can altcoins act as hedges or safe-havens for bitcoin?'' \emph{Finance Research Letters}, vol.~52, p. 103360, 2023.

\bibitem{mongardini2025midsummer}
A.~M. Mongardini and A.~Mei, ``A midsummer meme's dream: Investigating market manipulations in the meme coin ecosystem,'' \emph{arXiv preprint arXiv:2507.01963}, 2025.

\bibitem{krause2025meme}
D.~Krause, ``Meme coins and the trump effect: Deregulation, speculation, and the future of political cryptocurrencies,'' \emph{Speculation, and the Future of Political Cryptocurrencies}, 2025.

\bibitem{li2025trust}
Y.~Li, N.~Yao, Y.~Huo, and W.~Cai, ``Trust dynamics and bot-driven responses: An approach to rug pulls in solana meme coin markets,'' in \emph{Proceedings of the 17th ACM Web Science Conference 2025}, 2025, pp. 106--116.

\bibitem{la2023doge}
M.~La~Morgia, A.~Mei, F.~Sassi, and J.~Stefa, ``The doge of wall street: Analysis and detection of pump and dump cryptocurrency manipulations,'' \emph{ACM Transactions on Internet Technology}, vol.~23, no.~1, pp. 1--28, 2023.

\bibitem{tandon2021can}
C.~Tandon, S.~Revankar, and S.~S. Parihar, ``How can we predict the impact of the social media messages on the value of cryptocurrency? insights from big data analytics,'' \emph{International Journal of Information Management Data Insights}, vol.~1, no.~2, p. 100035, 2021.

\bibitem{lansiaux2022community}
E.~Lansiaux, N.~Tchagaspanian, and J.~Forget, ``Community impact on a cryptocurrency: Twitter comparison example between dogecoin and litecoin,'' \emph{Frontiers in Blockchain}, vol.~5, p. 829865, 2022.

\bibitem{mitman2025into}
T.~Mitman and J.~Denham, ``Into the meme stream: The value and spectacle of internet memes,'' \emph{new media \& society}, vol.~27, no.~6, pp. 3470--3486, 2025.

\bibitem{brichta2023fanning}
M.~Brichta, ``Fanning money: the cultural economy and participatory politics of dogecoin,'' \emph{International Journal of Communication}, vol.~17, p.~19, 2023.

\bibitem{serada2023happier}
A.~Serada, ``Happier than ever: The role of public sentiment in cryptocurrencies, meme stocks, and nfts,'' in \emph{Activist retail investors and the future of financial markets}.\hskip 1em plus 0.5em minus 0.4em\relax Routledge, 2023, pp. 35--53.

\bibitem{yogarajah2022hodling}
Y.~Yogarajah, ``‘hodling’on: Memetic storytelling and digital folklore within a cryptocurrency world,'' \emph{Economy and Society}, vol.~51, no.~3, pp. 467--488, 2022.

\bibitem{anton2024moon}
E.~Anton, M.~Aptyka, T.~D. Oesterreich, and F.~Teuteberg, ``To the moon with dogecoin! disentangling the causalities behind extrinsic and intrinsic motivations for memecoin investments,'' \emph{Journal of Decision Systems}, pp. 1--35, 2024.

\bibitem{philander2023meme}
K.~S. Philander, ``Meme asset wagering: Perceptions of risk, overconfidence, and gambling problems,'' \emph{Addictive Behaviors}, vol. 137, p. 107532, 2023.

\bibitem{long2025bridging}
H.-W. Long, N.-M. Wong, and W.~Cai, ``Bridging culture and finance: A multimodal analysis of memecoins in the web3 ecosystem,'' in \emph{Companion Proceedings of the ACM on Web Conference 2025}, 2025, pp. 1158--1161.

\bibitem{nobanee2023we}
H.~Nobanee and N.~O.~D. Ellili, ``What do we know about meme stocks? a bibliometric and systematic review, current streams, developments, and directions for future research,'' \emph{International Review of Economics \& Finance}, vol.~85, pp. 589--602, 2023.

\bibitem{d2022deep}
V.~D’Amato, S.~Levantesi, and G.~Piscopo, ``Deep learning in predicting cryptocurrency volatility,'' \emph{Physica A: Statistical Mechanics and its Applications}, vol. 596, p. 127158, 2022.

\bibitem{matsumoto2012some}
A.~Matsumoto, U.~Merlone, and F.~Szidarovszky, ``Some notes on applying the herfindahl--hirschman index,'' \emph{Applied Economics Letters}, vol.~19, no.~2, pp. 181--184, 2012.

\bibitem{coinmarketcap}
\BIBentryALTinterwordspacing
{CoinMarketCap}. (2025) Coinmarketcap. Accessed: Sep. 21, 2025. [Online]. Available: \url{https://coinmarketcap.com}
\BIBentrySTDinterwordspacing

\bibitem{etherscan}
\BIBentryALTinterwordspacing
{Etherscan}. (2025) Etherscan blockchain explorer. Accessed: Sep. 21, 2025. [Online]. Available: \url{https://etherscan.io}
\BIBentrySTDinterwordspacing

\bibitem{coincarp}
\BIBentryALTinterwordspacing
{CoinCarp}. (2025) Coincarp cryptocurrency data. Accessed: Sep. 21, 2025. [Online]. Available: \url{https://www.coincarp.com}
\BIBentrySTDinterwordspacing

\end{thebibliography}

\end{document}